\documentclass[a4paper,pre,twocolumn,floatfix]{revtex4-1}
\usepackage{graphicx}
\usepackage{amsmath}
\usepackage{amsthm}
\usepackage{amssymb}
\usepackage{natbib}

\begin{document}

\title{Revisiting the cavity-method threshold for random $3$-SAT}

\date{\today} 

\author{P. H. Lundow} 
\email{per.hakan.lundow@math.umu.se} 

\author{K. Markstr\"om}
\email{klas.markstrom@math.umu.se} 

\affiliation{Department of mathematics and mathematical statistics,
  Ume\aa{} University, SE-901 87 Ume\aa, Sweden}

\begin{abstract}
  A detailed Monte Carlo-study of the satisfiability threshold for
  random 3-SAT has been undertaken. In combination with a monotonicity
  assumption we find that the threshold for random 3-SAT satisfies
  $\alpha_3 \leq 4.262$.  If the assumption is correct, this means
  that the actual threshold value for $k=3$ is lower than that given
  by the cavity method.  In contrast the latter has recently been
  shown to give the correct value for large $k$. Our result thus
  indicate that there are distinct behaviours for $k$ above and below
  some critical $k_c$, and the cavity method may provide a correct
  mean-field picture for the range above $k_c$.
\end{abstract}


\maketitle

\theoremstyle{plain}
\newtheorem*{conjecture}{Conjecture}

\maketitle

%----------------------------------------------------------------------
\section{Introduction}
The properties of random $k$-SAT formulae has become one of the most
studied intersection points of computer science, mathematics and
physics. In this problem we have $n$ Boolean variables $x_i$ and we
construct a random Conjunctive Normal Form (CNF) formula $F$ by
picking $m$ clauses of size $k$ at random. Here each clause is the
disjunction, "OR", of $k$ literals, and each literal is either a
variable or its negation, leading to $2^k {n \choose k}$ possible
clauses.  The formula $F$ is satisfiable if there is an assignment of
values to the $x_i$:s such that every clause in $F$ becomes True.  If
$m$ is small then a random formula is with high probability
satisfiable and if $m$ is sufficiently large the formula is with high
probability not satisfiable.  In particular, it is believed, but not
known, that there exists constants $\alpha_k$ such that for a fixed
$\alpha = \frac{m}{n}$ less than $\alpha_k$ the probability for
satisfiability goes to 1 as $n$ grows, and for $\alpha$ larger than
$\alpha_k$ it goes to 0.  It is known that there exists some
$\alpha_k(n)$ such that this is true \cite{F}, but that the
$\alpha_k(n)$ is converging to a constant is only known for $k=2$, see
e.g., Ref.~\cite{ChR}, where $\alpha_2=1$, and sufficiently large
fixed $k$ \cite{DSS}.  Using methods from the theory of spin-glasses
the values of $\alpha_k$, and its existence as a constant, has been
calculated non-rigorously \cite{a,b}, and the results of
Ref.~\cite{DSS} show that this prediction for $\alpha_k$ is correct
for large enough $k$.

It has also been observed empirically that random CNFs with $\alpha$
close to $\alpha_k$ are harder to solve (find a satisfying assignment
for or refute) than when $\alpha$ is further away from $\alpha_k$. It
has repeatedly been speculated that this peak in the hardness of the
formulae is related to the clustering properties of the set of
solutions, as a function of $\alpha$. However, here there are no
corresponding rigorous hardness results, and since it is now known
that polynomial time solvable problems like random XOR-SAT have the
same type of clustering \cite{cl1,cl2} this connection is no longer
thought be straightforward.  The solution clustering in itself has
been verified for large $k$ \cite{Ac}. Another early product of
applying the cavity method to random $k$-SAT is the survey-propagation
algorithm.  This algorithm empirically demonstrated a good ability to
find solutions to satisfiable random $k$-SAT instances close to the
satisfiability threshold and it was conjectured that it would work for
all densities up to the threshold, unlike other randomized algorithms
which are known to fail before reaching the threshold.  However, this
has now been rigorously proven to not be the case, both for the
simpler belief-propagation method \cite{CO} and the full
survey-propagation method \cite{hetterich:LIPIcs:2016:6219}.  In
\cite{CO} the reason for this is discussed in detail, and one of the
reasons is that the cavity method makes some too simple assumptions on
the correlations in the model, for densities close to the threshold.

Since the existence of $\alpha_k$ has been established for large $k$,
and the related threshold is understood in quite some detail for
$k=2$, our aim has been to provide an improved test of the prediction
for $k=3$. Before the predictions from the cavity method arrived
several sampling studies of the thresholds were made, for many values
of $k$, but after the predictions were made no large scale study of
these predictions has been undertaken.  One obvious reason for this is
that the computer time needed for such studies grows exponentially
with the number of variables, and in order to get the required
accuracy a large number of samples is needed.  The latter is
especially important since many of the scalings used to analyse the
data in the earlier simulation papers were later ruled out by rigorous
mathematical results \cite{W}, thereby invalidating the method behind
those results.

We have sampled the random 3-SAT problem both with more variables than
in earlier studies, up to $n=375$ and a far larger number of samples per
density. In earlier papers typically a few thousand samples were used,
while for most values of $n$ we have several millions instead.  Our
main aim has been to provide an upper bound on the value of $\alpha_3$
and under a mild monotonicity assumption we find an upper bound of
$\alpha_3 \leq 4.262$.  This value is clearly smaller than the
cavity-method prediction $\alpha_*=4.26675$ \cite{b}, but closer to
the earlier \cite{AI} simulation estimate which arrived at $4.258$,
using an invalid scaling. It has already been noted \cite{KMR} that in
terms of the solutions space geometry the case $k=3$ differs from
$k\geq 4$, indicating that small values of $k$ might be exceptional,
and we will discuss possible reasons for the deviation of the
numerical prediction $\alpha_*$ from the actual value.

%-------------------------------------------------------------
\section{Sampling details}
In order to estimate the 3-SAT threshold we have sampled the random
3-SAT model for $n=4$, $8$, $16$, $32$ and $n=25, 50, \ldots, 375$. We
also attempted sampling for larger $n$ but there the sampling was so
slow that we could not generate the amount of data needed in order to
control the sampling noise. We used the MiniSAT solver to generate our
data \cite{Een}.   For each value of $n$ we produced random formulae 
with a fixed number of clauses $m$, for a range of values of $m$.

The number of samples were as follows, for $n=100, \ldots, 200$ we
have $N=4\times 10^6$ samples, for $n=225, \ldots, 300$, $N=10^6$, for
$n=325$, $N=5\times 10^5$, for $n=350$, $N=10^5$, and for $n=375$,
$N=1.4 \times 10^4$.  In each case we used densities in the interval
$[4.2,4.3]$.  For $n=350$ and $n=375$ we attempted to compensate for
the smaller number of samples by slightly increasing the number of
densities, but as we will see these two cases would still require more
samples in order to give sharp results.

We also sampled 2-SAT and 4-SAT, for $k=2$ we collected $10^4$ samples
for each size and density, and for $k=4$ we collected at least $10^4$
for each size and density for $n=50$, $75$, $100$, $125$.  For 2-SAT
we also used a data set produced by David Wilson \cite{W2}. This has
$10^4$ samples per size for $n=2^t$ where $t=1,\ldots, 20$.  The data 
from Wilson is produced in a different way from our own samples. Wilson 
starts with an empty formula $F_0$ and step by step produces a new formula 
$F_t$ from $F_{t-1}$ by adding a random clause to $F_{t-1}$, stopping 
when $F_t$ is unsatisfiable. The random formula $F_t$ is distributed in 
exactly the same way as a random formula with $n$ variables and $t$ clauses.  
For $k=2$ this sampling method is efficient due to the existence of a linear 
time algoritm for 2-SAT,  but for larger $k$  the standard method, which we 
have used, is more efficient.

%---------------------------------------------------------------------------
\section{The threshold for random 3-SAT}\label{3sat}
In order to estimate the value of $\alpha_3$ we have focused on the
value $\alpha(n,p)$ where the probability of being satisfiable is
equal to $p$, and in particular $p=\frac{1}{2}$.  The sharp threshold
result of Ref.~\cite{F} shows that, if the limit $\alpha_3$ exists,
the value $\alpha(n,p)$ will converge to $\alpha_3$ for any fixed
value $p$. However, the rate of convergence may depend on $p$.

The quantity $\alpha(n,1/2)$ has been used in several earlier studies,
e.g., Refs.~\cite{SK,AI}, where the approach has been to fit a
function of the form $a n+ b n^\beta$ to the estimated values of
$\alpha(n,1/2)$ for some range of values of $n$. In Ref.~\cite{AI} the
value $\beta=-2/3$ was found to give a good fit to the data.  However,
in Ref.~\cite{W} it was proven that there can exist at most one value
$p$ such that $\alpha(n,p)=\alpha_3 +o(n^{-\frac{1}{2}})$ and as
pointed out in Ref.~\cite{W} the experimental data indicates that the
unique such value for $p$, if it exists at all, is not
$\frac{1}{2}$. Hence a data fit of the type used in Refs.~\cite{SK,AI}
is unlikely to be valid, and if we change the value of $p$ by any
amount it is guaranteed that the form of the fitted function is valid
for at most one of the two values for $p$, no matter how small the
difference between them are.

\begin{figure}
  \begin{center}
    \includegraphics[width=0.48\textwidth]{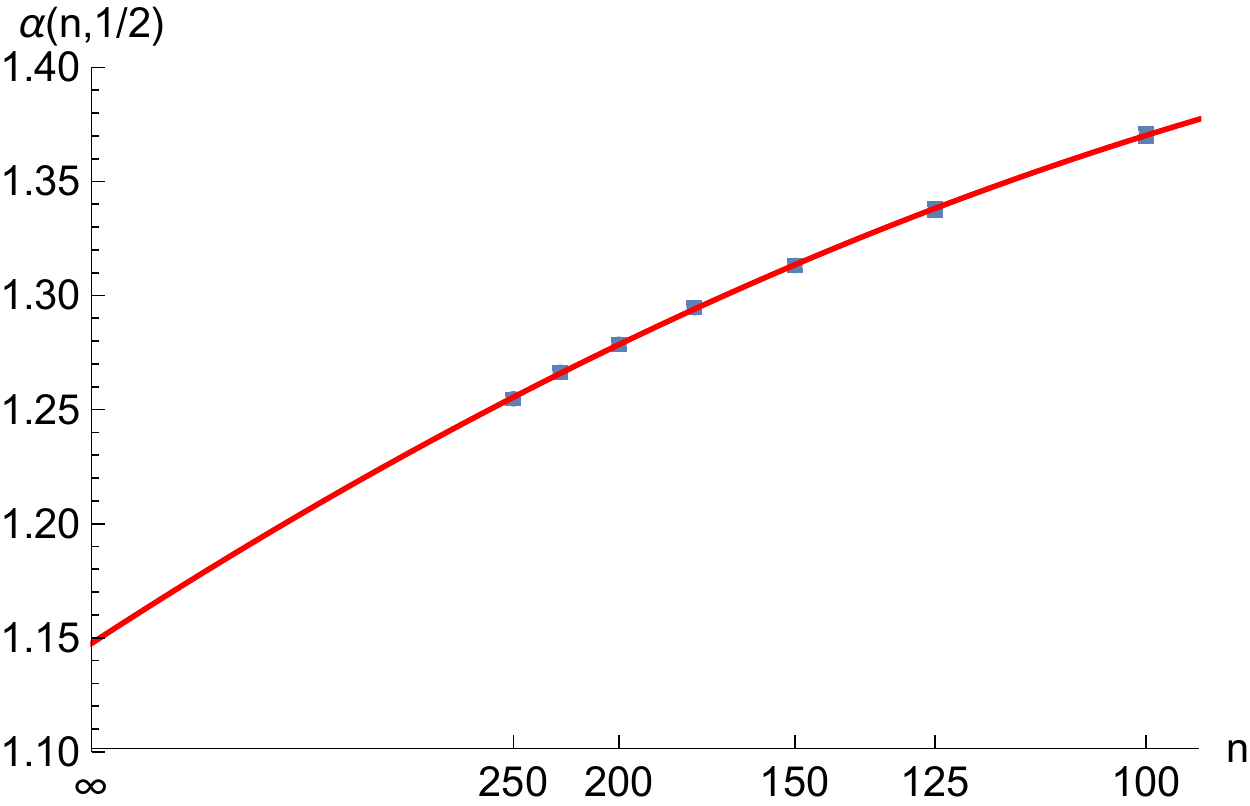}
  \end{center}
  \caption{(Colour on-line) For $k=2$, $\alpha(n,1/2)$ versus $n$ for
    $n=100, 125,\ldots, 250$ and the fitted polynomial (red curve)
    $1.15 + 30.1 x - 785 x^2$, where $x=1/n$. Error bars are
    smaller than the points.}\label{plotk2smallf}
\end{figure}

\begin{figure}
  \begin{center}
    \includegraphics[width=0.48\textwidth]{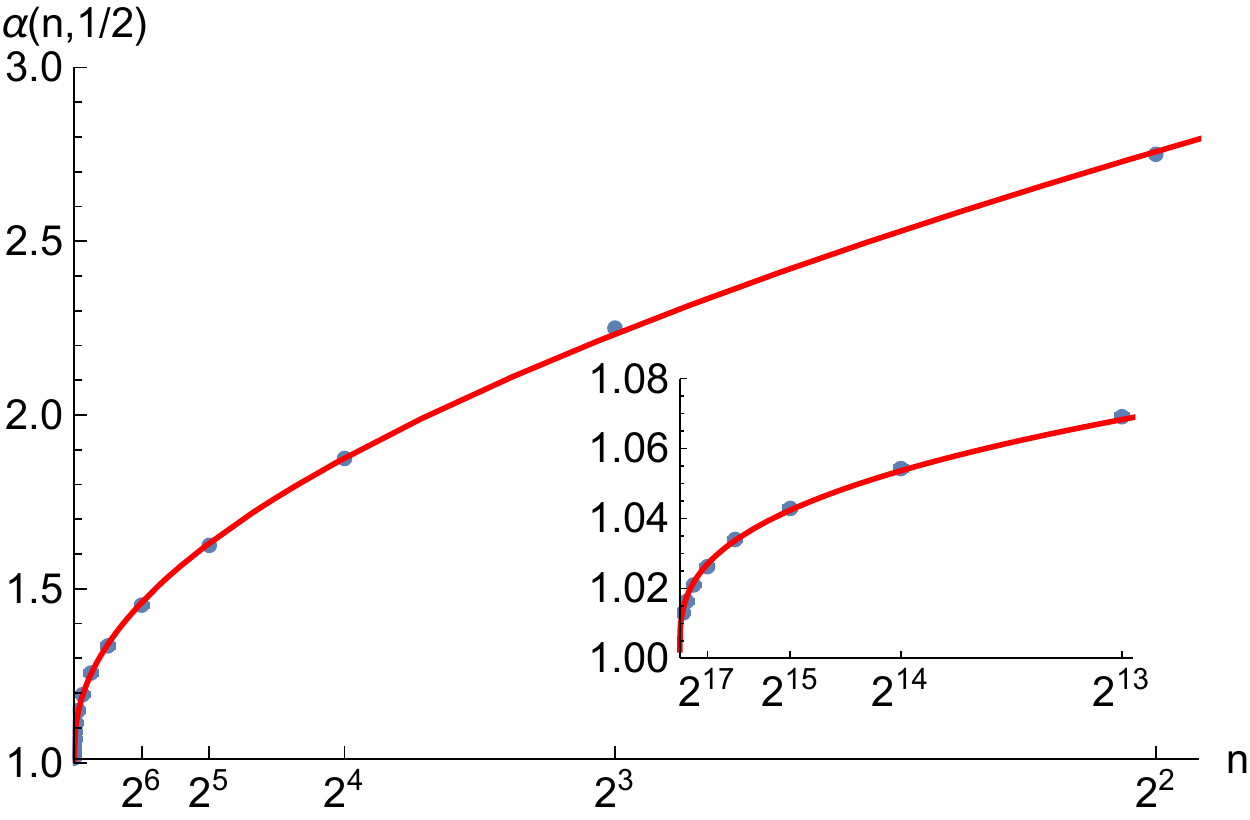}
  \end{center}
  \caption{(Colour on-line) For $k=2$, $\alpha(n,1/2)$ versus $n$ for
    $n=2^2, 2^3,\ldots, 2^{20}$ and the fitted curve (red) $1.00 +
    1.21 x^{1/3} + 2.51 x^{2/3}$, where
    $x=1/n$. Error bars are smaller than the points.}\label{plotk2largef}
\end{figure}

In order to demonstrate the discussed problem we look at the case
$k=2$, where we both have data for extremely large $n$ \cite{W2} and
rigorous results \cite{B1} on the threshold.  In
Fig.~\ref{plotk2smallf} we see the estimated values of $\alpha(n,1/2)$
as a function of $1/n$ for a range of $n$ similar to that used for
$k=3$. This graph was produced in the way which we will discuss for
3-SAT in the next section.  Here we know that $\alpha=1$ and the
scaling exponent for $1-\alpha(n,1/2)$ is $1/3$ \cite{B1}.
Nonetheless even a simple second degree polynomial gives a reasonable
fit to the data for $n\leq 250$.  Next, in Fig.~\ref{plotk2largef} we
see the same quantity but now for $n$ from $4$ up to $2^{20}$ and with
a fitted function based on the correct scaling exponent.  Here the
value of $\alpha(n,1/2)$ was produced by finding the median stopping
time in Wilson's data.  The median stopping time is identical to the
number of clauses $m$ given by $\lceil n\alpha(n,1/2)\rceil$, so the
two methods give easily comparable data.  The good fit of the
polynomial in the first figure is entirely due to the small values of
$n$ and has nothing to do with the correct asymptotics.  So, for the
case $k=2$ one can clearly be misled by small values of $n$.

\begin{figure}
  \begin{center}
    \includegraphics[width=0.48\textwidth]{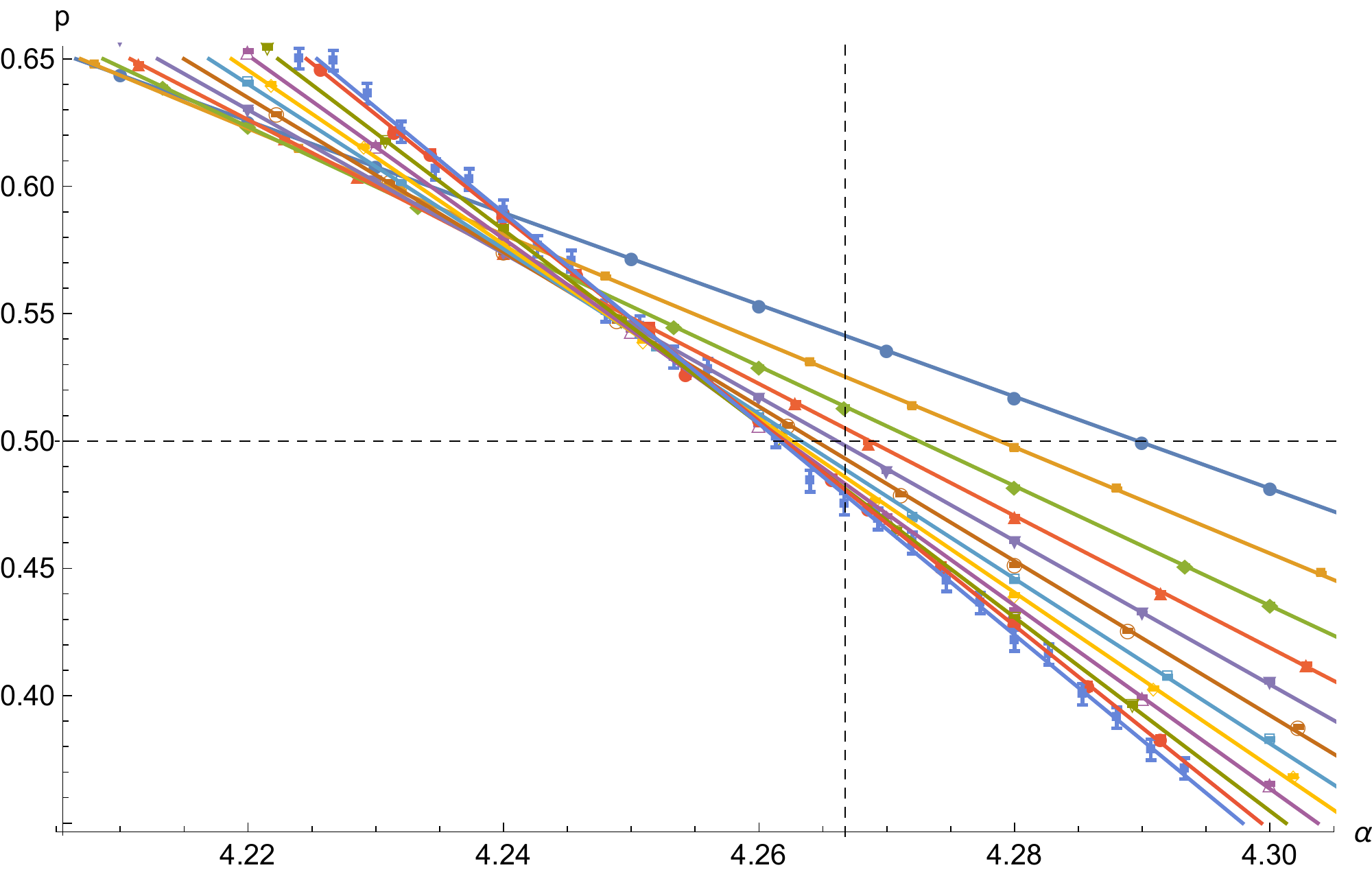}
  \end{center}
  \caption{(Colour on-line) For $k=3$, the probability $p$ of being
    satisfiable versus $\alpha$, together with fitted lines for
    $n=100,\ldots, 375$ (downwards at $\alpha=4.30$). Dashed lines at
    $\alpha=\alpha_*$ (vertical) and $p=1/2$ (horisontal).}\label{pl1}
\end{figure}

We now proceed to our data for $k=3$. In order to estimate the value
of $\alpha(n,1/2)$ we fitted, for each $n$, a line to the interval
where the probability $p$ for being satisfiable is in the range $|p
-\frac{1}{2}|\leq 0.15$, and then found the point where this line was
equal to $1/2$, using this as our estimate for $\alpha(n,1/2)$.  We
also tried polynomials rather than lines but in this interval the
curve is so close to linear that higher degree polynomials provided no
discernible improvement.  In Fig.~\ref{pl1} we see the sampled data
for the larger $n$ together with the fitted lines.

\begin{figure}
  \begin{center}
    \includegraphics[width=0.48\textwidth]{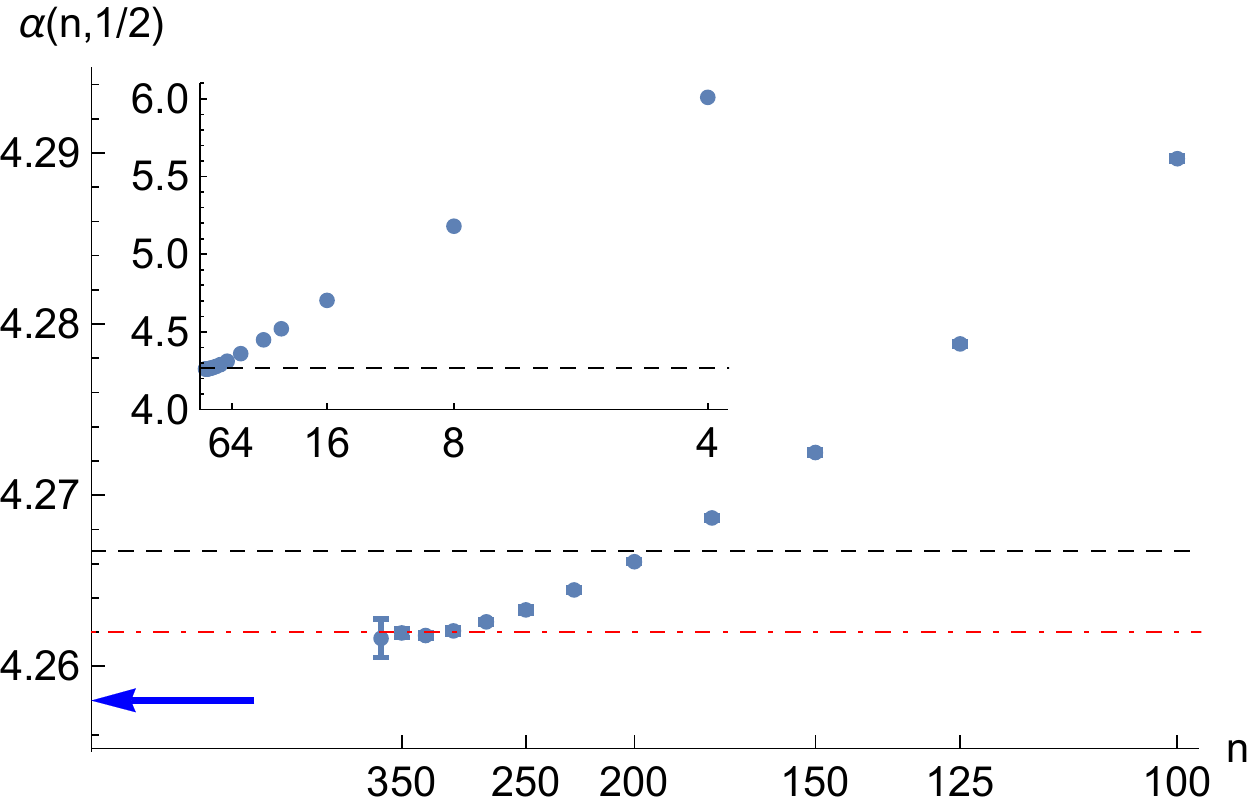}
  \end{center}
  \caption{(Colour on-line) For $k=3$, $\alpha(n,1/2)$ versus $n$ for
    $n=100, 125, \ldots, 375$. Black dashed line at
    $\alpha_*=4.26675$, red dot-dashed line at our upper bound
    $4.262$, blue arrow at estimate $4.258$ \cite{AI}. Inset shows
    zoomed-out version, for $n\ge 4$, dashed line at
    $\alpha_*$.}\label{plotk3}
\end{figure}

We estimate $\alpha(n, 1/2)$ for $n=100, 125,\ldots, 375$ as,
respectively,
\begin{eqnarray*}
  4.2897, 4.2788, 4.2725, 4.2687, \\
  4.2661, 4.2645, 4.2633, 4.2626, \\
  4.2621, 4.2618, 4.2619, 4.2616
\end{eqnarray*}
We have considered three sources for errors in these estimates, the
sampling noise, the degree of the polynomial fitted to the data, and
the choice of density values used in the fit.  The dominant error
turns out to be the sampling noise.  Since we have perfectly
independent samples we can do a correct error estimate for the
estimate by using bootstrap in the form of resampling, i.e., obtaining
estimates on different subsets of the data and finding the standard
deviation of the estimate under resampling.  All $n\leq 325$ give
similar values for the error estimate and in each case it is at most
$0.000176$, for $n=350$ we get $0.00027$ and for $n=375$ we get
$0.0011$. In Fig. \ref{plotk3} the error bars give the exact error
estimate for each $n$. As expected the size of the error closely
follows the number of samples.

We also considered the stability under using a polynomial of higher
degree than 1 in the fit to the data.  Using polynomials up to degree
4 this error turns out to be smaller than the sampling error, and is
in fact decreasing with $n$, indicating that the curve becomes more
and more linear in the given interval as $n$ grows.  We saw a similar
behavior when we used different subsets of the density values in the
fit, here the error for $n=375$ was less than 1\% of the sampling
error.
 
The $\alpha(n,1/2)$-values are shown in Fig.~\ref{plotk3}. Again we
see an almost linear behavior for small $n$, as the inset picture
shows, and then for the largest $n$ the points seem to level out. For
the last two points noise becomes noticable due to the too small
number of samples for those $n$.  The points in Fig.~\ref{plotk3} can
be well approximated by a second degree polynomial, but, as mentioned
before, from Ref.~\cite{W} we know that this is not a valid
scaling. In fact, we would expect the curve to behave as a suitable
root of $1/n$, just like in Fig.~\ref{plotk2smallf}, but we clearly do
not have large enough values of $n$ here to see the range where the
asymptotic behavior becomes dominant.

\begin{figure}
  \begin{center}
    \includegraphics[width=0.48\textwidth]{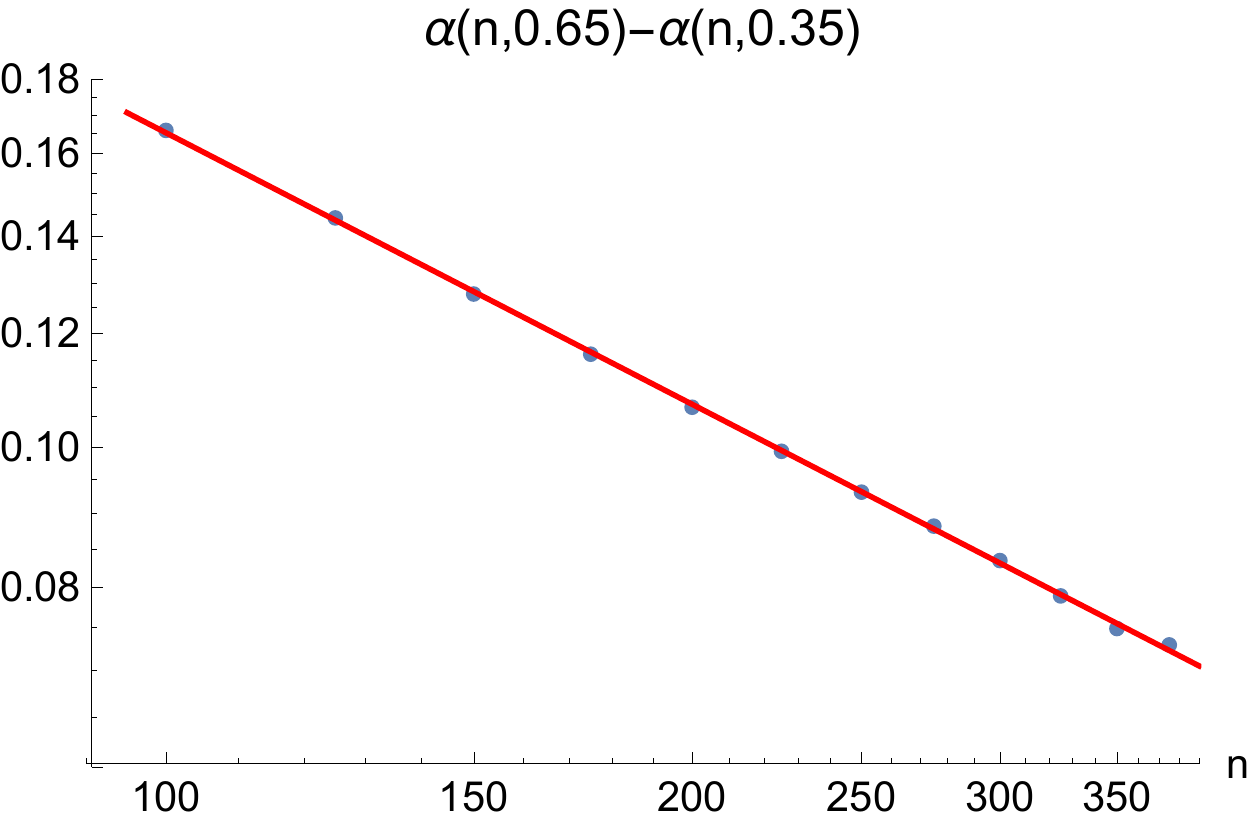}
  \end{center}
  \caption{For $k=3$, log-log plot of $\alpha(n,0.65)-\alpha(n,0.35)$
    versus $n$ for $n=100, 125,\ldots, 375$ and the fitted line (red)
    $1.0802 - 0.6255 x$, with $x=\ln n$.}\label{diff3}
\end{figure}

We find further evidence for the fact that we have too small values of
$n$ if we look at the width of the scaling window.  If we look at
$\alpha(n,0.65)-\alpha(n,0.35)$ we know from Ref.~\cite{W} that this
width cannot be $o(n^{-1/2})$, but in a log-log plot of this, as shown
in Fig.~\ref{diff3}, we see that we get the fitted line $1.0802 -
0.6255 x$. This gives a scaling of $n^{-0.625}$, which is ruled out
\cite{W}. The exponent $0.625$ is smaller than the $2/3$ found in
Refs.~\cite{SK,AI}. This could be due to the larger values of $n$ used
here and might indicate that we are at least getting closer to the
size range where the asymptotic scaling becomes visible.

With this in mind we find that one cannot give a credible estimate for
$\alpha_3$ with any accuracy based on this range of $n$ and have
instead taken the more modest aim of providing an upper bound on
$\alpha_3$.  In order to do this we have taken as our working
assumption that $\alpha(n,1/2)$ is in fact monotone in $n$, something
we believe to be true for large enough $n$.
\begin{conjecture}\label{mon}
  For any $k\geq 2$ there exists an $n_0$ such that for $n\geq n_0$
  the value of $\alpha(n,1/2)$ is decreasing in $n$.
\end{conjecture}
There are several reasons for believing that this type of monotonicity
should hold, and that $n_0$ will also be small. On one hand this is a
common occurrence for probabilistic combinatorial problems, and it is
also seen in many coupon-collector problems.

A coupon-collector problem has some base set $X$ and at each time step
$i$ a random subset $Y_i$ of $X$ is chosen with replacement, according
to some distribution for the $Y_i$, until all elements of $X$ are
covered by at least one $Y_i$.  That random $k$-SAT can be viewed as a
coupon-collector problem is a folklore result and has been used in
several published papers, e.g., Refs.~\cite{Zito, Kap}. Here the base
set $X$ is the hypercube $Q_n$ consisting of all binary strings of
length $n$, and each random set $Y_i$ is a random subcube of dimension
$n-k$, corresponding to the solutions ruled out by a clause of size
$k$.  A $k$-SAT formula is unsatisfiable if the corresponding
collection of sets $Y_i$ cover all elements of the hypercube $Q_n$.
As mentioned in connection with Wilson's data the value
$\alpha(n,1/2)$ corresponds exactly to the median stopping time of the
coupon-collector process.  For the simplest coupon-collector problem
the median, as well as the full distribution of the stopping time, was
derived in Ref.~\cite{ErRe}, and after normalization to make it
converge, it does indeed decrease to it's asymptotic value.  General
coupon-collector problems have been studied, e.g., in Ref.~\cite{FLM}
where $k$-SAT is also discussed, and for many such examples the type
of monotonicity conjectured above can be proven. In fact we know of no
natural examples where this type of monotonicity is known to fail, but
there is no general monotonicity result which includes the case of
$k$-SAT for fixed $k$.

A second reason for expecting both monotonicity and a low value of
$n_0$ comes from the seminal results of Ref.~\cite{CS}.  There a
rigorous analysis of the structure of a random unsatisfiable $k$-SAT
formula $F$ was undertaken for all densities $\alpha$, not only for
values above the threshold. One of the main results is that there
exists a function $g_k(\alpha)$ such that the smallest unsatisfiable
sub-formula of $F$ has at least $n g_k(\alpha)$ variables, and this
function $g_k(\alpha)$ is decreasing with $\alpha$.  So, the
unsatisfiability of $F$ is explained by the appearance of an
unsatisfiable sub-formula $F'$ which has linear size, but the relative
size is smaller for higher densities $\alpha$.  However, the set of
unsatisfiable formulae on $n g_k(\alpha)$ variables is more restricted
for small $n$ than for larger $n$, since there are more ways of
realizing such a formula for larger $n$, and likewise is more
restricted the larger $g_k(\alpha)$ is.  Hence one should expect the
set of such formulae to be closer to its asymptotic behavior for small
values of $g_k(\alpha)$, i.e, for large densities $\alpha$.  This
would then mean that for larger densities we see a faster convergence
to the asymptotic probability of satiability, also indicating that
$\alpha(n,1/2)$ should move to the left.  Indeed, the point
$\alpha(n,1/2)$ also corresponds to the density where the median
number of unsatisfiable sub-formulae in $F$ is at least 1, and if we
add as little as $O(\ln(n))$ the expected number of unsatisfiable
sub-formulae of size at least $n g_k(\alpha)$ in $F$ will be at least
polynomial in $n$.

For $k=2$ the conjecture agrees with both data, as shown in
Fig.~\ref{plotk2largef}, and with what one would expect from the
mathematical results \cite{ChR,B1}, even though this is not explicitly
proven in the latter.  Our sequence of values for $\alpha(n,1/2)$ is
compatible with this assumption with the exception for the value at
$n=350$, but a closer examination of the data for the two largest
values of $n$ shows that those estimates are too noisy for the needed
accuracy.  Under the monotonicity assumption and a very pessimistic
view of the sampling errors we can then confidently give the bound
$$\alpha_3 \leq 4.2620.$$ In Fig.~\ref{plotk3} we see values of
$\alpha(n,1/2)$ for $n\geq 100$ together with lines indicating the
cavity-method prediction $\alpha_*=4.26675$, our asymptotic upper
bound, and an arrow marking the early estimate $4.258$ from
Ref.~\cite{AI}.

%-------------------------------------------------------------------
\section{Discussion}
As we have seen, our upper bound for $\alpha_3$ is incompatible with
the cavity-method prediction from Refs.~\cite{a,b}. We note that our
estimate for $\alpha(200,1/2)$ is already below the predicted
asymptotic value, that is in the range for $n$ where we have
$N=4\times 10^6$ samples per density so we are confident that our
estimate is accurate.

\begin{figure}
  \begin{center}
    \includegraphics[width=0.48\textwidth]{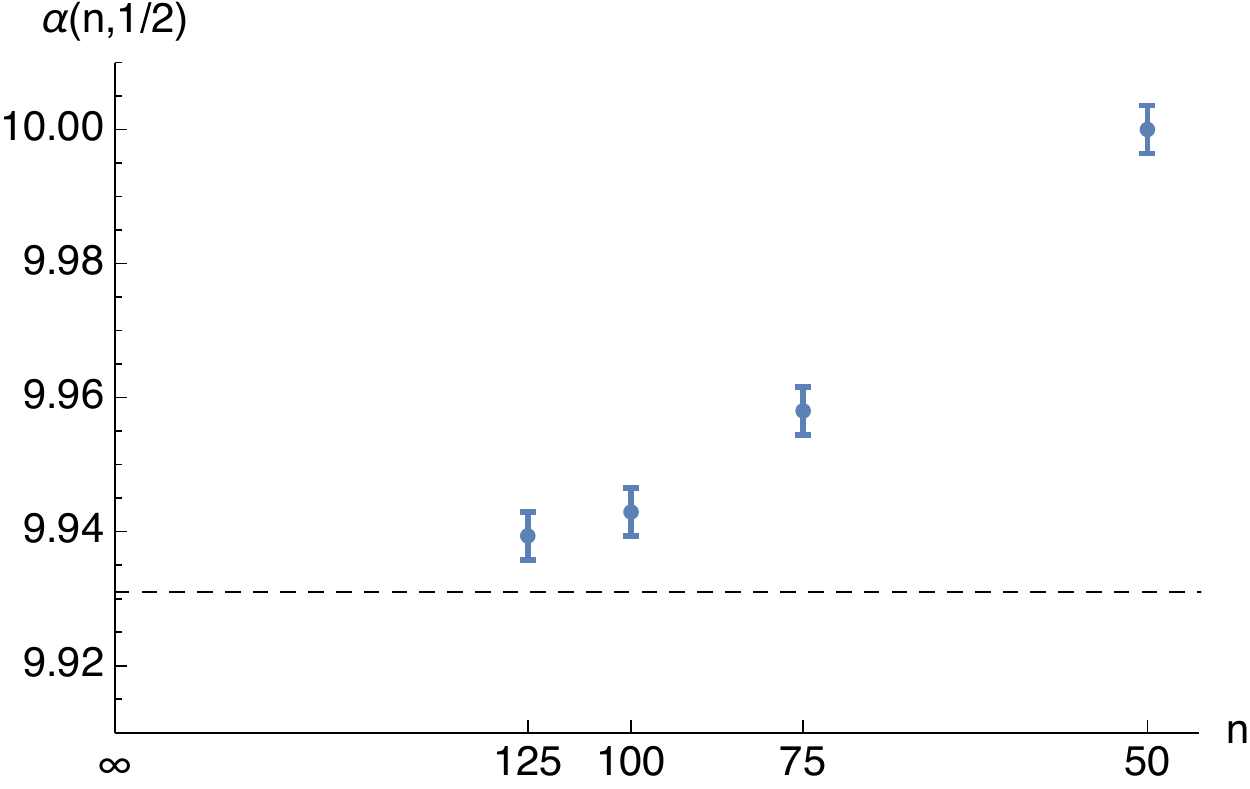}
  \end{center}
  \caption{For $k=4$, $\alpha(n,1/2)$ versus $n$ for $n=50$, $75$,
    $100$, $125$. Dashed line at cavity-method estimate
    $\alpha_*=9.931$.}\label{plotk4}
\end{figure}

One explanation for the contradiction between our bound and $\alpha_*$
could of course be that $\alpha(n,1/2)$ is not monotone, but this
would require a strong, and in our opinion surprising, finite-size
correction to the observed behavior, which would also differ from
what we see at $k=2$.  In Fig.~\ref{plotk4} we show a plot of
$\alpha(n,1/2)$ for $k=4$, for $n=50$, $75$, $100$, $125$ and we once
again see a monotone decrease with $n$.  Here the values (estimated to
$10.000$, $9.962$, $9.945$, $9.941$, respectively) stay above the
cavity-method prediction $9.931$ for $k=4$, but the values of $n$ are
even smaller than for $k=3$.

Another explanation could lie in the numerical determination of
$\alpha_*$ in Ref.~\cite{b}.  In that paper a set of equations for
$\alpha_*$ is derived, but they are in terms of an optimum over a set
of distribution functions which are not explicitly known.  In order to
find $\alpha_*$ they perform a numerical search over a quite
complicated search space and it is possible that this search has in
fact not found a correct optimum. In an earlier paper \cite{a} the
smaller value $4.256$ was stated, but then changed \cite{b} after it
was found that the numerical procedure was sensitive to the type of
random number generator used in the search. However, this problem
would have to be unusually sensitive if numerical errors has led to
incorrect optima both above and below the actual value.

The third and perhaps most intriguing possibility is that the cavity
method itself, as used in Refs.~\cite{a,b}, does in fact not give a
correct prediction for $\alpha_3$.  We know from Ref.~\cite{DSS} that
the cavity method does give the correct value for $\alpha_k$ for large
enough $k$, but those authors have stated that they do not think that
their proof can be extended all the way down to $k=3$.  It has also
been found~\cite{KMR} that the cavity method predicts that other
thresholds, which describe properties of the set of solutions to a
satisfiable $k$-SAT formula, behave differently for $k=3$ and $k\geq
4$. In the former case some of the generally distinct thresholds
coincide. Those authors also found that the analysis of the method
would require changes for $k=3$, thus indicating that for the cavity
method itself the case $k=3$ is distinct.

In combination with our results this leads to a picture where the
cavity method may provide the correct mean-field type behavior for
$k$ above some critical $k_c$, leaving a few distinct cases for lower
$k$, much in analogy with the high and low-dimensional behavior for
classical phase transitions, like random walks, percolation and the
Ising model.  In either of the two latter cases the well known
prediction $\alpha_*=4.26675$ is not correct and a further
investigation of the case $k=3$ for random $k$-SAT seems worthwhile,
both from a mathematical and a physical point of view.

%-------------------------------------------------------------------
\begin{acknowledgments}
  We would like to thank the anonymous referee for constructive
  criticism on the first version of our manuscript.  The simulations
  were performed on resources provided by the Swedish National
  Infrastructure for Computing (SNIC) at High Performance Computing
  Center North (HPC2N).  This work was supported by the Swedish
  strategic research programme eSSENCE.  This work was supported by
  The Swedish Research Council grant 2014--4897.
\end{acknowledgments}

%----------------------------------------------------------------
%\bibliographystyle{apsrev}
%\bibliography{papers}

\end{document}